\documentclass[a4paper]{jpconf}
\usepackage{graphicx}
\usepackage{amssymb}
\begin{document}
\rightline{DAMTP-2009-90}
\rightline{UG-09-69}
\rightline{MIT-CTP-4096}
\vskip -1truecm
\title{On massive gravitons in  2+1 dimensions}

\author{Eric Bergshoeff${}^1$, Olaf Hohm${}^2$ and Paul Townsend${}^3$}

\address{${}^1\,$Centre for Theoretical Physics, University of Groningen, Nijenborgh 4, 9747 AG Groningen, The Netherlands}

\address{${}^2\,$Center for Theoretical Physics, Massachusetts Institute of Technology, Cambridge, MA 02139, USA}

\address{${}^3\,$Department of Applied Mathematical Physics and Theoretical Physics, Centre for Mathematical Sciences, University of Cambridge, Wilberforce Road, Cambridge, CB3 0WA, U.K.}

\ead{e.a.bergshoeff@rug.nl; ohohm@mit.edu; P.K.Townsend@damtp.cam.ac.uk}

\begin{abstract}
The Fierz-Pauli (FP) free field theory for massive spin-$2$ particles can be extended, in a spacetime of (1+2) dimensions (3D), to a  generally covariant parity-preserving interacting field theory, in at least two ways. One is  ``new massive gravity'' (NMG),  with an action that involves curvature-squared terms. Another is 3D ``bigravity'', which  involves non-linear couplings of the FP tensor field to 3D Einstein-Hilbert gravity.
We review the proof  of the linearized equivalence of both ``massive 3D gravity'' theories to FP theory,  and we comment on their similarities and differences.

\end{abstract}

\section{Introduction: Fierz-Pauli and NMG}\setcounter{equation}{0}
%\medskip

There has been increased interest in recent years in the possibility of massive gravitons. This  is motivated in part by the discovery of cosmic acceleration, which might be explainable in terms of an infra-red modification of general relativity that gives the graviton a small mass (e.g. \cite{Eckhardt:2009kq}), and in part by the idea that some theory involving massive gravitons could be the low energy limit of a  non-critical string-theory underlying QCD (e.g. \cite{'tHooft:2007bf}). As for so many other gravitational physics issues, it may be useful to consider the possibilities for massive gravitons in the simpler context of a three-dimensional (3D) spacetime.

The standard free-field description of massive gravitons, i.e. massive particles of spin-$2$, is that due to Fierz and Pauli (FP) in terms of a symmetric tensor field, which we will call $f_{\mu\nu}$. The FP action is
\begin{equation}\label{FPact}
S_{FP}[f] = S_{EH}^{(2)}[f] - \frac{1}{4} m^2 \int d^3 x\, \left(f^{\mu\nu}f_{\mu\nu} -f^2\right)\, ,
\end{equation}
where $f= \eta^{\mu\nu} f_{\mu\nu}$ for Minkowski metric $\eta$, which we assume to have ``mostly plus'' signature, and
\begin{equation}
S_{EH}^{(2)}[f] = -\frac{1}{2}\int d^3x \, f^{\mu\nu} G_{\mu\nu}^{(lin)}(f)\, ,
\end{equation}
where $G_{\mu\nu}^{(lin)}(f)$ is the linearized Einstein tensor expressed as a self-adjoint  differential operator acting on the tensor field $f$; the FP action therefore reduces to the quadratic approximation to the Einstein-Hilbert (EH) action in the massless limit.  The field equations are equivalent to
\begin{equation}\label{FPeqs}
\left(\Box -m^2\right) f_{\mu\nu}=0\, , \qquad \partial^\mu f_{\mu\nu}=0\, , \qquad  f=0\, .
\end{equation}
The differential subsidiary condition is needed to remove ghost modes and the tracefree condition is needed to remove lower spin modes.
In 4D the FP equations propagate the 5 helicity states of a massive spin 2 particle. In 3D they propagate the two $\pm2$ helicity states of a massive spin 2 particle.

The main issue of interest  is whether one can generalize the FP free field theory to a consistent interacting theory. One could try to couple
$f_{\mu\nu}$ to some linear combination of the  stress tensor  and its trace but this leads to difficulties (see \cite{Goldhaber:2008xy} for a recent review). In 3D there is another option. To see this, we first solve the differential subsidiary condition by writing
\begin{equation}
f_{\mu\nu} = -\frac{1}{2} \epsilon_{\mu} {}^{\eta\rho}\epsilon_{\nu} {}^{\tau\sigma}\partial_\eta\partial_\tau  \, h_{\rho\sigma} \equiv G_{\mu\nu}^{\rm (lin)}(h)
\end{equation}
for some symmetric tensor potential $h$. The remaining FP equations become
\begin{equation}
\left(\Box -m^2\right) G^{\rm (lin)}_{\mu\nu}(h)=0\, , \qquad R^{\rm (lin)}(h) =0\, ,
\end{equation}
where $R^{\rm (lin)}(h)$ is the trace of the linearized Ricci tensor $R^{\rm (lin)}_{\mu\nu}(h)$.
These equations are derivable from the quadratic action
\begin{equation}\label{NMGquadact}
S_{NMG}^{(2)}[h] =  \int d^3x\, \left\{ \frac{1}{2} h^{\mu\nu} G_{\mu\nu}^{\rm (lin)}(h) +
\frac{1}{m^2} G^{\mu\nu}_{\rm (lin)}(h)S^{\rm (lin)}_{\mu\nu}(h)  \right\}\, ,
\end{equation}
where
\begin{equation}
S^{\rm (lin)}_{\mu\nu}(h)  = R^{\rm (lin)}_{\mu\nu}(h) - \frac{1}{4} \eta_{\mu\nu} R^{\rm (lin)}\, .
\end{equation}
This new quadratic action for symmetric tensor potential $h$ has an obvious extension to a non-linear generally covariant  action  for metric  $g_{\mu\nu}=\eta_{\mu\nu} + h_{\mu\nu}$; this yields ``new massive gravity''  (NMG) \cite{Bergshoeff:2009hq}.

It is convenient to give the NMG action in a slightly more general  form involving a dimensionless parameter $\sigma$; in appropriate units, this action is \cite{Bergshoeff:2009hq}
\begin{equation}\label{original}
S_{NMG}[g] = \int d^3x\,  \sqrt{-\det g} \left[ \sigma R + m^{-2} K \right] \, ,
\end{equation}
where
\begin{equation}\label{Kinv}
K := R^{\mu\nu}R_{\mu\nu} - \frac{3}{8} R^2 \ = \  G^{\mu\nu} S_{\mu\nu}\, , \qquad  S_{\mu\nu} := R_{\mu\nu} - \frac{1}{4}g_{\mu\nu} R\, .
\end{equation}
The tensor $S_{\mu\nu}$ is the 3D Schouten tensor, which plays an important role in conformal gravity as the gauge potential for conformal boosts\footnote{This is true in any dimension but the expression for the Schouten tensor in terms of the Ricci tensor is dimension dependent. In $D$ spacetime dimensions it is defined as $S_{\mu\nu} = (D-2)^{-1} R_{\mu\nu} - [2(D-1)(D-2)]^{-1} R \, g_{\mu\nu}$.}.  To recover the quadratic action (\ref{NMGquadact}) we should set $\sigma=-1$, which means that the EH term has the `wrong' sign (relative to the EH action of general relativity).

We have arrived at the action for NMG by a method \cite{Andringa:2009yc} that guarantees its on-shell equivalence to the FP theory in the linearized limit. This method is a very general one \cite{Bergshoeff:2009tb} but it does not guarantee {\it off-shell} equivalence.
However, this can be proved very simply for NMG by means of the  following alternative  action  involving an `auxiliary' tensor field $f$  \cite{Bergshoeff:2009hq}:
\begin{equation}\label{altact}
S_{NMG}[g,f] = \sigma S_{EH}[g] +  \int d^3x \sqrt{-\det g} \left[  f^{\mu\nu} G_{\mu\nu} - \frac{1}{2} m^2
g^{\mu[\rho} g^{\nu]\sigma} f_{\mu\nu}f_{\rho\sigma} \right]\, .
\end{equation}
This action, which is only 2nd order in derivatives,  is quadratic in the tensor  $f_{\mu\nu}$. The $f$ field equation  is $f_{\mu\nu} =(2/m^2)S_{\mu\nu}$; using this equation to eliminate $f_{\mu\nu}$ (or integrating it out by Gaussian functional integration in the path integral)  one recovers the action (\ref{original}) in terms of the metric alone.  This is true for any value of $\sigma$, so for $\sigma=0$ the action
(\ref{altact}) constitutes a 2nd order version of the  ``pure-$K$'' model studied in  \cite{Deser:2009hb}. 
Focusing  on NMG proper, we set  $\sigma=-1$. To linearize about the Minkowski vacuum,  we  now write
\begin{equation}\label{gexpansion}
g_{\mu\nu} = \eta_{\mu\nu} + h_{\mu\nu} - f_{\mu\nu}
\end{equation}
and expand the action in powers of the Minkowski space tensors $(h,f)$. To quadratic order we find that
\begin{equation}\label{NMGquad}
S^{(2)}_{NMG}[h,f] = - S_{EH}^{(2)}[h] + S_{FP}[f]\, .
\end{equation}
Since the quadratic EH action for $h$ propagates no modes, the modes propagated  by the linearized NMG theory are the same as those of the FP theory\footnote{In \cite{Bergshoeff:2009hq} we set $g_{\mu\nu}=\eta_{\mu\nu}+h_{\mu\nu}$ and used the $h$ equation to eliminate $h$ from the action. This step was criticized in \cite{Deser:2009hb} because it involves a back-substitution that is not generally permissible. However, the legitimacy of this back-substitution is manifest in  the new basis implicit in  (\ref{gexpansion}) because it just eliminates the quadratic EH term, 
which propagates no modes. This change of basis was first made in \cite{Nakasone:2009bn}; it  had occurred independently to us and we used it  in our general analysis of  unitarity for the  `cosmological' extension of NMG \cite{Bergshoeff:2009aq}.}.

Higher-derivative extensions of 4D general relativity have been investigated since Weyl's 1918 attempt at a unified theory of gravity and electromagnetism. A special feature of curvature-squared invariants is that they contribute to the quadratic approximation in an expansion about  a Minkowski vacuum,  so they contribute to the linearized field equations, which become fourth order rather than second order. Equations of higher than second order  generally propagate ghosts (modes of negative energy) in addition to physical modes, and this is what happens here, in 4D:  in addition to the massless graviton, there is also a massive scalar mode and a massive spin-$2$ ghost \cite{Stelle:1977ry}.
The 3D case is  different because the EH action by itself does not propagate any modes, so the massless graviton is now absent. This means that we are free to change the overall sign of the action to arrange for the massive spin-$2$ modes to be physical, i.e. not ghosts;  we can then ensure that they are not tachyons by a choice of  relative sign for the EH and curvature-squared terms. However, there is still the massive scalar mode, which is now a ghost. Remarkably,  the mass of this scalar ghost mode goes to infinity as the relative coefficient of the  two independent curvature-squared terms is tuned to the value that yields the `$K$-combination' of  (\ref{Kinv}), so we end up with the unitary massive pure spin-$2$ model that is NMG.  All of this can be read off  from  the results of Nishino and Rajpoot \cite{Nishino:2006cb} for the general 3D model with EH and curvature-squared invariants but these authors considered only the `right-sign' EH term and (presumably as a consequence) did not draw attention to the special features  of the $K$-combination.

One motivation for considering the EH action with the `wrong' sign is that this is  a feature of ``topologically massive gravity'' (TMG) \cite{Deser:1981wh}, which  is also ``higher-derivative''   because it involves the third order, and parity violating,  Lorentz Cherns-Simons (LCS) term. TMG  propagates only a single spin 2 mode, which is possible in 3D  if parity is violated. 
This possibility can also be realized by a first-order ``self-dual'' spin-$2$ theory that can be thought of as the ``square-root'' of the 3D FP theory \cite{Aragone:1986hm}.  If the differential subsidiary condition of this model is solved, the remaining equations become precisely those of linearized TMG, so TMG can be deduced by the same procedure that we used above to arrive at NMG. Moreover, in this case the on-shell equivalence extends immediately to an off-shell equivalence because with only one propagating mode one can always choose the overall sign of the action so as to ensure that it is physical rather than a ghost.  Unitarity for NMG is a more delicate matter because there are two propagating modes. Unitarity follows  from the proof of off-shell equivalence to the FP theory reviewed above, but it can also be checked explicitly via a canonical analysis \cite{Deser:2009hb}. A nice feature of this  method is that one sees explicitly the cancellation of the higher time derivatives; in other words, the action is actually second order in time derivatives despite being  fourth order in space derivatives. This is reminiscent of Ho\v rava gravity \cite{Horava:2009uw} but without any  violation of Lorentz invariance!

There is a natural generalization of NMG that is suggested by TMG: one simply adds a LCS term to the action. This yields what
we have called ``general massive gravity'' (GMG) \cite{Bergshoeff:2009hq}. There are now two mass parameters, which can be traded for the two masses $m_\pm$ of the two spin-$2$ modes of helicities $\pm2$, which we may assume to be such that $m_-\ge m_+ >0$.  As parity flips the sign of helicities, a model with $m_+\ne m_-$ breaks parity. When $m_+=m_-$ the parity-violating LCS term is absent and we recover NMG. The limit $m_-\to\infty$ for fixed $m_+$ yields TMG, in which the curvature-squared term is absent,  so both NMG and TMG are special cases of GMG. Conversely, it is possible to recover GMG by a ``soldering'' of two TMG models \cite{Dalmazi:2009es}, one propagating a helicity +2 mode with mass $m_+$ and the other propagating a helicity $-2$ mode with mass $m_-$.

\section{3D bigravity}\label{sec:bg}
%\medskip

In the case of 4D massless gravitons, it is known that any theory of interacting gravitons must reduce to general relativity at low energy \cite{Weinberg:1965rz,Deser:1969wk,Boulware:1974sr}. Even if these low energy theorems continue to apply in 3D,  they
cannot restrict the interactions of massive gravitons because these will  decouple at sufficiently low energy. There is therefore no obvious reason why  NMG should be the unique interacting extension of the 3D FP theory. Indeed, an alternative  parity-preserving generally covariant massive gravity model was recently proposed by Ba\~nados and Theisen \cite{Banados:2009it}; their ``bigravity'' model for massive 3D gravitons is a 3D version of the  4D ``$f$-$g$ gravity'' of  Isham {\it et al.}  \cite{Isham:1971gm}.

The  Ba\~nados-Theisen (BT) model involves a dimensionless parameter, which we shall call $\alpha$, and two tensor fields; we
shall use a different basis for these two fields and we call them $(\tilde g,f)$. Also, we omit here the cosmological terms included in  \cite{Banados:2009it}; we will later discuss briefly the cosmological extension.  For $\alpha(\alpha+1)> 0$, the BT action is
\begin{equation}\label{bgaction}
S_{BT}[\tilde g,f] =\alpha S_{EH}[\tilde g+ \beta f] + S_{EH}[\tilde g -\alpha\beta f]  -
\int d^3 x\ {\cal U} (\tilde g,f)
\end{equation}
where
\begin{equation}\label{alphaeq}
\beta =1/\sqrt{\alpha(\alpha+1)}\, .
\end{equation}
The last term in the action is the potential term, defined by some choice of  scalar density ${\cal U}$ constructed from  the metric tensor and the $f$-tensor, without derivatives. The choice made in \cite{Banados:2009it} has the form
\begin{equation}
U(\tilde g,f) =  \frac{1}{2} m^2\sqrt{-\det \tilde g}\  \tilde g^{\mu[\rho} \tilde g^{\nu]\sigma} f_{\mu\nu}f_{\rho\sigma}
+ {\cal O}\left(f^3\right)\, .
\end{equation}
A feature of the BT action is that it assumes invertibility of both $\tilde g$ and $\tilde g + \beta f$. We can interpret the  invertibility constraint on, say, $\tilde g$ as the geometrical condition that $\tilde g$ be a metric tensor but  invertibility of $\tilde g + \beta f$ then puts a constraint on the tensor field $f$ that is not so obviously geometrical.

The proof that  linearized bigravity is equivalent to the FP theory is simple. We write
\begin{equation}
\tilde g_{\mu\nu} = \eta_{\mu\nu} + h_{\mu\nu}
\end{equation}
and expand the action (\ref{bgaction}) in powers of the Minkowski space tensors $(h,f)$. The result to quadratic order is
\begin{equation}\label{BTquad}
S^{(2)}_{BT}[h,f] = \left(\alpha +1 \right)  S^{(2)}_{EH}[h] + S_{FP}[f]\, .
\end{equation}
This is equivalent to the FP theory for any $\alpha$, because the quadratic EH action for $h$ propagates no modes, although we should recall the constraint $\alpha(\alpha+1)>0$.

If $\beta$ is initially taken as a free parameter then one finds that  unitarity is violated if  $\alpha(\alpha+1)<0$ irrespective of the choice of $\beta$, and that (\ref{alphaeq}) can be imposed  without loss of generality when $\alpha(\alpha+1)>0$.
When $\alpha=0$ the action (\ref{bgaction}) is equivalent to the EH action, so we can ignore this case. When $\alpha=-1$, the action
(\ref{bgaction}) is trivial but it is also not equivalent to the action considered in \cite{Banados:2009it}; in this case we should
instead consider the action
\begin{equation}
\left. S_{BT}[\tilde g,f]\right|_{\alpha=-1} =- S_{EH}[g] + S_{EH}[g -f] -   \int d^3 x\ {\cal U} (\tilde g,f)\, .
\end{equation}
Linearization now yields the quadratic action
\begin{equation}\label{quadminusone}
\left. S^{(2)}_{BT}[h,f]\right|_{\alpha=-1} =  \int d^3 x\, \left\{ h^{\mu\nu} G_{\mu\nu}^{\rm (lin)}(f) - \frac{1}{4} m^2
\left( f^{\mu\nu}f_{\mu\nu} -f^2\right)\right\}\, ,
\end{equation}
but this is precisely what one gets from linearization of  (\ref{altact}) when $\sigma=0$. As observed in \cite{Bergshoeff:2009aq}, the $h$ tensor is now a Lagrange multiplier imposing a constraint that has the solution $f_{\mu\nu}= 2\partial_{(\mu} A_{\nu)}$. Using this in the $f$ equation, we find that
\begin{equation}
G_{\mu\nu}^{\rm (lin)}(h) = m^2 \left[\partial_{(\mu}A_{\nu)} - \eta_{\mu\nu}\left(\partial\cdot A\right)\right]\, .
\end{equation}
This equation determines the metric fluctuation tensor $h$ in terms of the source, which is itself constrained by the linearized Bianchi identity; this constraint is just the 3D Maxwell equation  $\partial^\mu F_{\mu\nu}=0$, where $F_{\mu\nu} = 2\partial_{[\mu}A_{\nu]}$. It is also true that the action (\ref{quadminusone}) reduces to the 3D Maxwell action upon making the substitution $f_{\mu\nu}= 2\partial_{(\mu} A_{\nu)}$. Either way, we see that a single massless mode is propagated,
of undefined spin because spin is not
defined for massless particles in 3D. This is consistent with
the canonical analysis of the  ``pure-K'' model \cite{Deser:2009hb}.

%The action then reduces to the Maxwell action
%\begin{equation}
%\left. S^{(2)}_{BT}[h,f]\right|_{\alpha=-1}  \to  -\frac{1}{4} m^2 \int d^3x F^{\mu\nu} F_{\mu\nu}\, , \qquad
%F_{\mu\nu} \equiv  \partial_\mu A_\nu - \partial_\nu A_\mu\, .
%\end{equation}

\section{Cosmological NMG and 3D bigravity}\label{sec:vs}
%\setcounter{equation}{0}
%\medskip

We conclude with a few brief observations on the cosmological extension of NMG and 3D bigravity. For NMG one may add the cosmological term
\begin{equation}
S_{cos}[g] =  -2\lambda m^2 \int d^3 x \, \sqrt{-\det g}
\end{equation}
where $\lambda$ is a new dimensionless parameter. It is convenient in this context to also allow the parameter $\sigma$ to be arbitrary, and to allow for $m^2<0$. Anti-de Sitter (adS) vacua are possible for $\lambda\ne0$, and in such cases one may study the nature of the associated
2D conformal field theory (CFT) \cite{Brown:1986nw}. It turns out that this has a negative central charge whenever the quadratic approximation to the bulk theory (expanded about the adS vacuum) is unitary, and vice versa \cite{Bergshoeff:2009aq}.  There is one exception, for $\lambda=3$, in which the massive gravitons are replaced by massive spin 1 modes and the central charge of the CFT is zero; in this case the CFT is of logarithmic type \cite{Grumiller:2009sn}. 

In the bigravity case we may add the cosmological term
\begin{equation}
S_{cos}[\tilde g] = \frac{2\left(\alpha+1\right)}{\ell^2}  \int d^3 x \, \sqrt{-\det \tilde g}\, .
\end{equation}
This allows an adS vacuum of radius $\ell$ for the metric $\tilde g$, with $f=0$. Linearization about adS was
considered in \cite{Banados:2009it}, and also in \cite{Afshar:2009rg}. 
`Bulk' unitarity requires either that $\alpha\le-1$ or that $\alpha\ge0$.
{}From our understanding of the
results of \cite{Banados:2009it}, the central charge of the associated 2D CFT is (in our notation) a positive factor times $(\alpha+1)$, which would mean that it is negative for $\alpha<-1$ but positive for $\alpha>0$.
It would appear from this that 3D cosmological bigravity is analogous to cosmological NMG for $\alpha<-1$ but not for $\alpha>0$. This can be seen in vestigial form from the Minkowski space actions: when $\alpha<-1$ the
quadratic actions (\ref{BTquad}) are related by a rescaling (which is trivial at $\alpha=-2$) but not when $\alpha>0$ because of the different sign of the quadratic EH term; although this does not affect the conclusions concerning propagating modes, and may continue to be irrelevant  at the  non-linear level in a Minkowski vacuum,  it makes a significant difference in an adS vacuum.

In conclusion, both NMG and 3D bigravity  constitute generally covariant interacting extensions of the  free-field 3D Fierz-Pauli model for massive particles of spin 2. Each will likely have its advantages and disadvantages. Whether either will provide insight into the problem of massive 4D gravity remains to be seen.

\ack
EB thanks the organizers of the 2009 Spanish Relativity Meeting ERE2009 for a stimulating environment.
PKT is supported by an EPSRC Senior Fellowship.
The work of OH is supported by the DFG -- The German Science Foundation
and in part by funds provided by the U.S. Department of Energy (DoE) under
the cooperative research agreement DE-FG02-05ER41360.

%The command \verb"\ack" sets the acknowledgments heading as an unnumbered section.

\end{document}